\documentclass[
    ,final            
  ]
  {aipproc}

\layoutstyle{8x11double}

\usepackage{amssymb}

\begin{document}

\title{Mixing-induced CP violating sources for electroweak baryogenesis
from a semiclassical approach}


\author{Yu-Feng Zhou}{
  address={Korea Institute for Advanced Study, Seoul 130-722, Korea}
}



\begin{abstract}

The effects of flavor mixing in electroweak baryogenesis is investigated
in a generalized semiclassical WKB approach. Through calculating the
nonadiabatic corrections to the particle currents it is shown that
extra CP violation sources arise from the off-diagonal part of the
equation of motion of particles moving inside the bubble wall. The 
 mixing-induced source is of the first order in derivative
expansion of the Higgs condensate, but is oscillation suppressed.
The numerical importance of the mixing-induced source is discussed
in the Minimal Supersymmetric Standard Model and compared with the
source term induced by semiclassical force.
\end{abstract}

\maketitle

Electroweak baryogenesis (EWBG) is a promising scenario for explaining
baryon number asymmetry in the universe, unlike other scenarios valid
at grand unification scale it can be tested by the upcoming collider
experiments.
An efficient way to generate large baryon number asymmetry is to generate
it nonlocally, through a charge transportation mechanism \cite{Cohen1991}.
In some models such as MSSM and the two-Higgs-doublet model, the bubble
wall of the electroweak phase transition is typically thick \cite{Cline1998,Moreno1998}
for the particles giving dominant contributions, namely the particle
typical Compton wave length $\lambda\sim1/T$ is much shorter than
the wall width $L_{w}\sim(10-20)/T$ where $T$ is the critical temperature.
In this regime the validity of semiclassical approach \cite{Cline:2001rk,Cline:2000nw,Cline2000,Huber:2000mg,Joyce1994}
should be justified for single flavor case, which provides an intuitively
simple description by treating particle transportation as motion of
WKB wave packages. In a slowly moving and CP violating Higgs condensate
background, the dispersion relation for particles and antiparticles
are modified differently, contributing to different semiclassical
forces, which leads to a net excess or deficit of the particle number
which can be converted into left-handed fermion number asymmetry.
The asymmetry of the local fermion density then get transported in
front of the bubble wall, which bias the baryon number violating processes. 

The problem becomes more involved in the multiple flavor case, where
the CP violating mass matrix has nontrivial space-time dependence.
The CP violating effects may show up not only in the dispersion relations,
but also in the mixing of states.
In this work we generalize
the semiclassical method by taking into account the non-adiabatic
corrections from the spatially varying flavor mixings \cite{Zhou:2008yd}. We restrict
ourselves in the parameter region where the mixing between the local
mass eigenstates are relatively small, and can be treated as perturbations
to the equation of motion for local mass eigenstates.

In two flavor mixing case the explicit form of the equation array
for left-handed component can be rewritten as

\begin{equation}
\left(\begin{array}{cc}
D_{11} & D_{12}\\
D_{21} & D_{22}\end{array}\right)\left(\begin{array}{c}
L_{s1}\\
L_{s2}\end{array}\right)=0 \ , 
\label{eq:EOM-mass}\end{equation}
with\begin{eqnarray}
D_{11} & = & \omega^{2}+\partial_{z}^{2}-m_{1}^{2}+2\Sigma_{11}\partial_{z}+isA_{11}(\omega-is\partial_{z})
\nonumber\\
D_{12} & = & 2\Sigma_{12}\partial_{z}+isA_{12}(\omega-is\partial_{z})
\nonumber\\
D_{21} & = & 2\Sigma_{21}\partial_{z}+isA_{21}(\omega-is\partial_{z})
\\
D_{22} & = & \omega^{2}+\partial_{z}^{2}-m_{2}^{2}+2\Sigma_{22}\partial_{z}+isA_{22}(\omega-is\partial_{z})
\nonumber \ .
\end{eqnarray}
Since the off-diagonal elements $D_{12(21)}$ contain differential
operators, one can {\it not} obtain decoupled equations for $L_{s1(2)}$
separately.

From the dispersion relation one can deduce the
group velocity $v_{g}\equiv(\partial\omega/\partial p_{c})_{z}$ and
semiclassical force $F_{i}\equiv\omega(dv_{g}/dt)$ respectively
\begin{eqnarray}
v_{gi} & = & \frac{p_{0i}}{\omega}-s\frac{m_{i}^{2}\mbox{Im}A_{ii}}{2\omega^{2}p_{0i}}
\ , \nonumber\\
F_{i} & = & -\frac{m_{i}m_{i}'}{\omega}-s\frac{\left(m_{i}^{2}\mbox{Im}A_{ii}\right)'}{2\omega^{2}} \ ,
\label{eq:force}
\end{eqnarray}
where $\omega$ is the energy of the state, 
$\Sigma=U^\dagger \partial U$ and $A=U^\dagger M\partial M^{-1} \partial U$
with $M$ the mass matrix in flavor basis and $U$ is the rotation matrix
diagonalizing $M^\dagger M$.  $p_c$ is the canonical momentum and 
$p_0^2=\omega^2-m^2$. 
Note that only the second term in the force term is CP violating,
and is proportional to the spin of the local mass states.

Taking the off-diagonal terms $D_{12}$ and $D_{21}$ as perturbations,
the solutions can be written in a generic form 
\begin{equation}
L_{si}=L_{si}^{(0)}+L_{si}^{(1)} \ , 
\end{equation}
where $L_{si}^{(0)}$ is the lowest order solution satisfying
\begin{equation}
D_{ii}L_{i}^{(0)}=0 \ ,
\label{eq:LowestOrderEq}\end{equation}
and $L_{si}^{(1)}$ are the corrections due to the off-diagonal terms.
The lowest order solution $L_{si}^{(0)}$ is obtained by the usual
WKB wave ansatz

\[
L_{si}^{(0)}=w_{i}e^{i\int^{z}p_{ci}(z')dz'} \ ,
\]
where $p_{ci}$ is the canonical momentum and the function $w_{i}$
provides the correct normalization for $L_{si}^{(0)}$.

Substituting the off-diagonal terms in to the equation, the first
order perturbation takes the following form\begin{eqnarray}
L_{si} &= & L_{si}^{(0)}+L_{si}^{(1)} \simeq L_{si}^{(0)}+\epsilon_{i}L_{si}^{(0)}
\end{eqnarray}
which are mixtures of the two unperturbed states. 
The mixing parameter
for particle $i$ to the first order of derivative is given by

\begin{equation}
\epsilon_{i}=i\frac{2\Sigma_{ij}p_{0j}+A_{ij}(s\omega+p_{0j})}{m_{i}^{2}-m_{j}^{2}},
\end{equation}
The mixing parameter for particle $j$ can be simply obtained by
replacing $i\leftrightarrow j$ from the above expressions. It is
clear that the expansion is only valid for $\partial_{z}m_{i}/(\Delta m^{2})\ll1$. 
The momentum dependencies comes from the differentiation
operators in off-diagonal element $D_{12}$.

The CP violating force term, as it is proportional to spin $s$, only
contribute to the spin-weighted density. To facilitate the comparison
with the force term, we give the spin-weighted mixing-induced
source term
 \begin{eqnarray*}
S_{Li} & = & \sum_{s}\frac{s}{2}\partial_{\mu}j_{Li}^{s\mu}\\
 & = & \frac{2m_{i}m_{j}}{m_{i}^{2}-m_{j}^{2}}\left[\mbox{Im}\Sigma g_{L}(p_{0i},p_{0j})-\frac{m_{i}}{m_{j}}\mbox{Im}\Pi_{ij}g_{R}(p_{0i},p_{0j})\right]\\
 &  & \cdot(p_{0i}-p_{0j})\sin\int^{z}(p_{cj}-p_{ci})dz'  \ ,
\label{eq:source}
\end{eqnarray*}
where $\Pi$ is a similar quantity as $\Sigma$ but for right-handed field. The functions
$g_{L,R}$ are given by
\begin{eqnarray*}
g_{L,R}(p_{0i},p_{0j})&=&\frac{\omega\mp p_{0j}}{\sqrt{p_{0i}p_{0j}(\omega+p_{0i})(\omega+p_{0j})}}
\\
&&-\frac{\omega\pm p_{02}}{\sqrt{p_{0i}p_{0j}(\omega-p_{0i})(\omega-p_{0j})}}
\end{eqnarray*}
which is a momentum odd function.

Following the standard process to derive the transportation equations,
one arrives at the usual form of the diffusion equation
\begin{equation}
-k_{i}\left(D_{i}\xi_{i}^{''}+v_{w}\xi'\right)+\tilde{\Gamma}_{ik}^{d}\sum_{j}\xi_{j}^{(k)}\simeq S_{F}+S_{M}
\end{equation}
with $\tilde{\Gamma}_{ik}^{d}=k_{i}\Gamma_{ik}^{d}$ the interaction rate.

The mixing and force induced source terms have the following form after thermal-average 
\begin{equation}
S_{F}=-\frac{k_{i}v_{w}D_{i}}{T\left\langle \left(\frac{p_{z}}{\omega}\right)^{2}\right\rangle }\left\langle \frac{p_{z}}{\omega}\delta F_{zi}\right\rangle ',\mbox{ and }S_{M}=k_{i}\left\langle S_{i}^{pl}\right\rangle 
\label{source-term}
\end{equation}
where $\xi\equiv\mu_{i}/T$ the rescaled chemical potential and $D_{i}=\left\langle (p_{z}/\omega)^{2}\right\rangle /\Gamma_{i}^{t}$
the diffusion constant.

%
%
In MSSM, the chargino transportation provides a dominant CP violating
source to electroweak baryogenesis. The asymmetry in chargino number
is converted into the asymmetry in left-handed top quarks through
Yukawa interactions. The two Higgsino $SU(2)$ doublets are $\tilde{h}_{1}=(\tilde{h}_{1L}^{0},\tilde{h}_{1L}^{-})^{T}$
and $\tilde{h}_{2}=(\tilde{h}_{2L}^{+},\tilde{h}_{2L}^{0})^{T}$ respectively.
Together with the two charged gauginos $\tilde{W}_{L}^{+}$ and $\tilde{W}_{L}^{-}$,
the charginos are combined into two left- and right-handed four component
spinors as $\psi_{L}=(\tilde{W}_{L}^{+},\tilde{h}_{2L}^{+})^{T}$
and $\psi_{R}=((\tilde{W}_{L}^{-})^{c},(\tilde{h}_{1L}^{-})^{c})^{T}$.
The chargino mass term has the form $\bar{\psi}_{R}M(z)\psi_{L}$
in the wino-higgsino space with mass matrix \begin{equation}
M(z)=\left(\begin{array}{cc}
M_{2} & gH_{2}(z)\\
gH_{1}(z) & \mu\end{array}\right) \ ,
\end{equation}
where $M_{2}$ and $\mu$ are soft supersymmetry breaking parameters
containing CP phases and $H_{1}(z)$ and $H_{2}(z)$ the Higgs vacuum
expectation values (VEVs).

\begin{figure}
 \includegraphics[width=.23\textwidth]{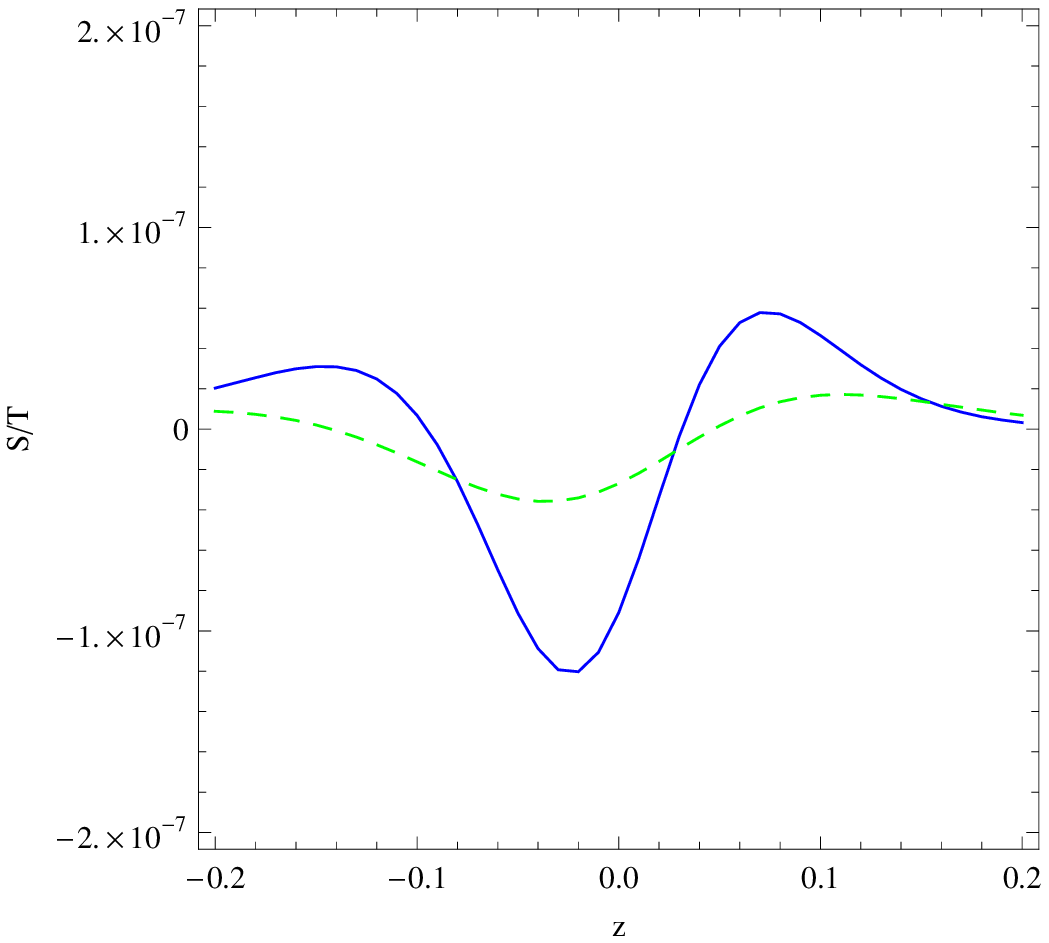}
 \includegraphics[width=.23\textwidth]{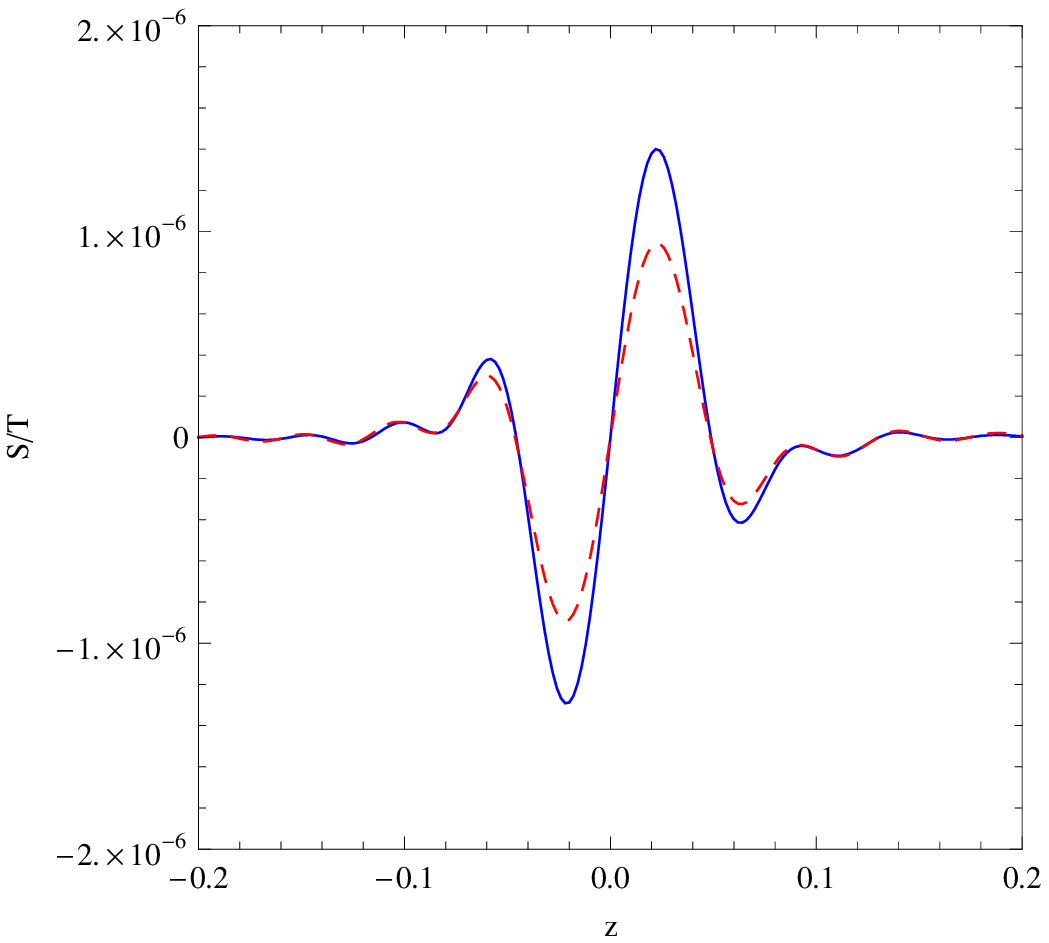}
\caption{(a)Left, the semiclassical force induced source term as a function
of $z$ ($\mbox{GeV}^{-1}$). (b) Right, the mixing-induced source
term as a function of $z$. The two curves in each plot corresponds
to $L_{w}=10/T$ (solid) and $15/T$ (dashed) respectively. The MSSM
parameters are fixed at $M_{2}=150$ GeV and $\mu=100$ GeV, with
$\phi_{\mu}=0.02$.}
\label{fig:1}
\end{figure}

\begin{figure}
\includegraphics[width=.23\textwidth]{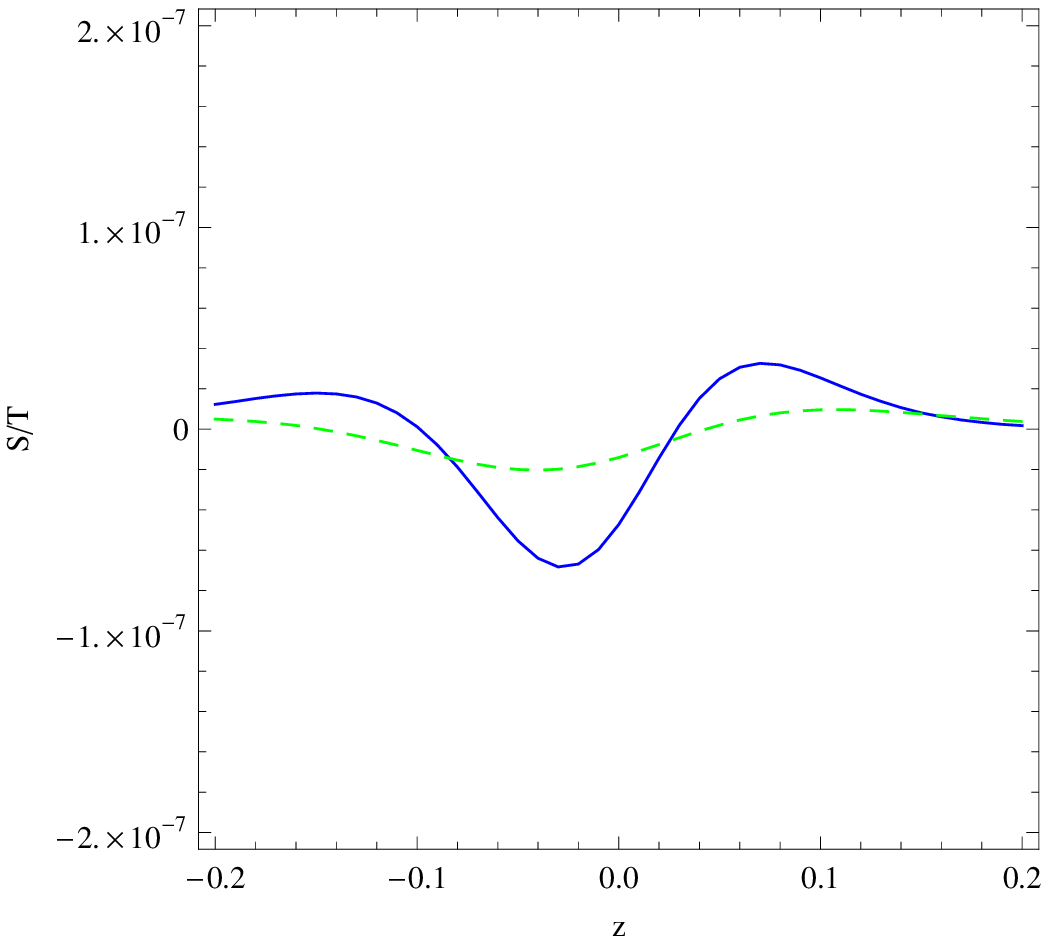}
 \includegraphics[width=.23\textwidth]{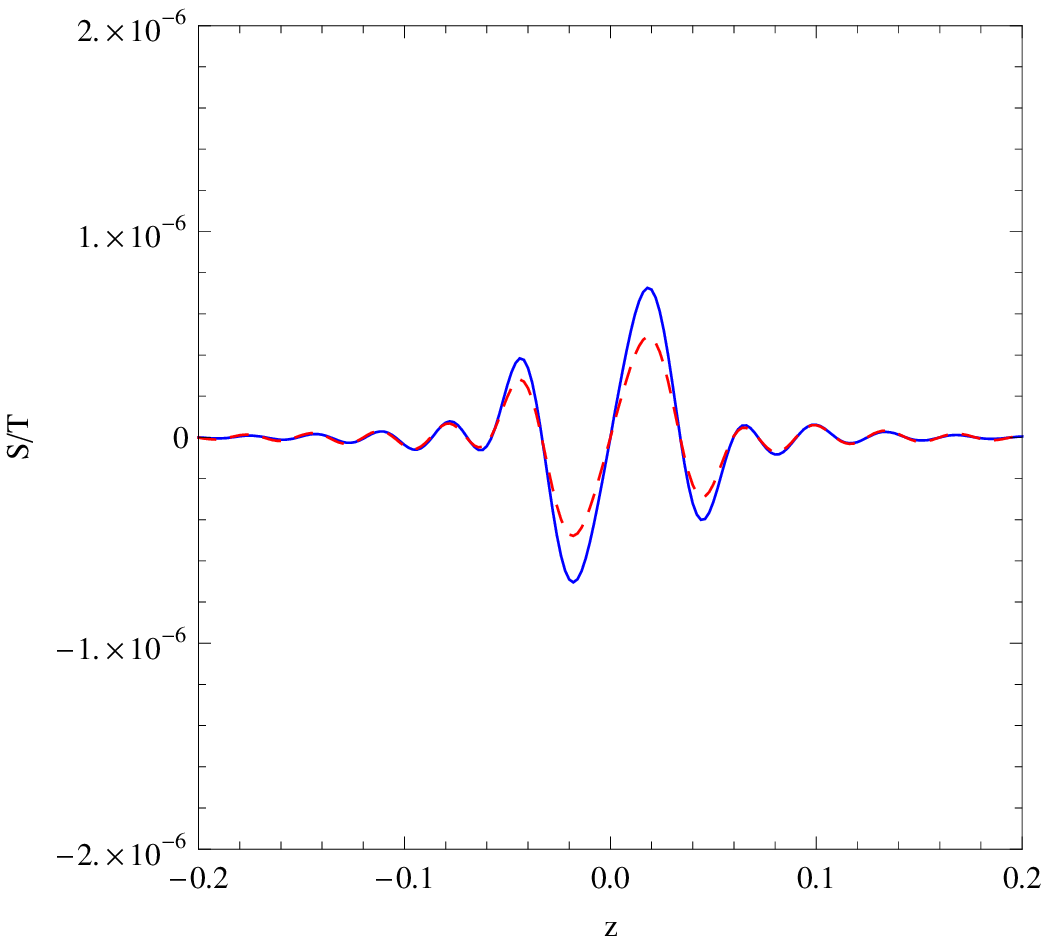}
\caption{ The same as Fig.\ref{fig:1} with $M_{2}=250$ GeV.
}
\label{fig:2}
\end{figure}

In Fig.\ref{fig:1}, we give the source term calculated from
Eq.(\ref{source-term}) and rescaled by $1/T$ from both semiclassical
force and mixing-induced source term for wall width $L_{w}=10/T$ and
$15/T$ respectively. The MSSM softbreaking parameters are fixed at
$M_{2}=200$GeV and $\left|\mu\right|=100$GeV corresponding to a
chargino mass differences $m_{1}^{}-m_{2}^{}=111$ GeV in the broken
phase. The CP phase $\phi_{\mu}$ is set to a typical value of
$\phi_{\mu}=0.02$. The curves given in
Fig.\ref{fig:1} show that both of source terms have nontrivial spatial
dependencies. However, their origin are quite different.  The variation of
the force term comes from the third derivative of the kink-type Higgs
condensate, which has one minimal and two maximums.  While the
mixing-induced source term varies due to both the wall profile
variation and the oscillation. The oscillation leads to multiple local
minimum appearing in the curve and is suppressed at large distance by
the wall profile. For $M_{2}$ around 200 GeV, the mixing induced
source term peaks at $z\simeq0.03$ with an amplitude
$S_{M}/T\simeq1.6\times10^{-6}$, much larger than that from
semiclassical force which peaks at $z\simeq0.08$ with
$S_{F}/T=0.6\times10^{-7}$. 
 In Fig.\ref{fig:2}, we give the results for a larger $M_{2}=250$
GeV, and still fix $\left|\mu\right|=100$GeV, corresponding to a
larger chargino mass differences $m_{1}-m_{2}^{}=157$ GeV in the
broken phase. One sees that for a larger chargino mass difference
both the source terms becomes smaller as they are $1/\Lambda$ suppressed.
The oscillation in the mixing-induced source term becomes obvious
and the wave length is shorter, thus the oscillation suppression is
stronger. The mixing induced source term peaks at $z\simeq0.03$ with
an amplitude $S_{M}/T\simeq8\times10^{-7}$. Although significantly
reduced, it still much larger than that from semiclassical force which
peaks at $z\simeq0.08$ with $S_{F}/T\simeq4\times10^{-8}$.

To estimate the oscillation suppression effects it is useful to define
an averaged source over the wall width
\begin{equation}
\bar{S}_{M(F)}\equiv\frac{1}{TL_{w}}\int_{0}^{L_{w}}S_{M(F)}(z)dz \ .
\end{equation}
We calculate the averaged source for different $M_{2}=200\sim500$
GeV and list the results in Tab.I  in Ref.\cite{Zhou:2008yd} The
results show  that
$\bar{S}_{M}$ dominates over $\bar{S}_{F}$ in the range $200\mbox{GeV}\lesssim M_{2}\lesssim350\mbox{GeV}$.
With the value of $M_{2}$ increasing, the averaged source term $\bar{S}_{M}$
drops rapidly. For $L_{w}=10/T$, at $M_{2}=350$ GeV, the mixing
induced source is only $5.4\%$ of that at $M_{2}=200$ GeV, the suppression
is due to the increased oscillation frequency. The suppression in
force term $\bar{S}_{F}$ is mainly from the factor $1/\Lambda$,
which makes it decrease slowly. At $M_{2}=350$ GeV, it is still about
$30\%$ as large as that at $M_{2}=200$ GeV. At $M_{2}=200$ GeV,
their relative size is $\bar{S}_{M}/\bar{S}_{F}=14.6$. For large
$M_{2}=350$ GeV, although they are close in size, the mixing source
still dominates with $\bar{S}_{M}/\bar{S}_{F}=2.63$. This dominance
has a mild dependence on the wall width. For $L_{w}=15/T$, the relative
size between the two kind of sources remains roughly the same, although
both of the source term becomes smaller. The mixing-induced source
(semiclassical force) term is $\sim40\%(\sim30\%)$ of that at $L_{w}=10/T$.
When the value of $M_{2}$ is around 450 GeV, the two type of source
term becomes comparable in size. For a very large $M_{2}=500$ GeV,
the semiclassical force term becomes dominate. The mixing-induced
source term is about an order of magnitude smaller. 

In conclusion, we have studied the effects of flavor mixing in a generalized
WKB approach in which the off-diagonal terms in the equation of motion
are taken into account as perturbations. With the presence of a slowly
moving CP violating bubble wall, an extra mixing-induced CP violating
source appears which exhibit an oscillation behavior in analogy to
the neutrino mixings. 
%
%
We have made a numerical study of the oscillation suppression effects
for chargino case in MSSM and shown that 
%
%
for a light $200\lesssim M_{2}\lesssim350$ GeV
even in the small mixing
case the mixing-induced source already indicate that a significant
enhancement of the final baryon number asymmetry is possible, which
will make MSSM a more realistic model for EWBG. 





\bibliographystyle{aipproc}   

\bibliography{ewvg2}

\end{document}